\documentclass[reprint, amsmath,amssymb, aps,]{revtex4-1} 	

\usepackage{graphicx}
\usepackage{dcolumn}								
\usepackage{bm}
\usepackage{bm,amsmath}
\usepackage{braket}
\newcommand{\average}[1]{\ensuremath{\langle#1\rangle} } 

\newcommand{\up}{\uparrow}	
\newcommand{\dn}{\downarrow}										

\usepackage{color}
\begin{document}

\preprint{APS/123-QED}

\title{Weak antilocalization in spin-orbit coupled lattice systems: effect of non-adiabatic transitions and estimation of spin relaxation length}

\author{Hiroshi Hayasaka}
 \email{HAYASAKA.Hiroshi@nims.go.jp}
\affiliation{%
 Elements Strategy Initiative Center for Magnetic Materials, National Institute for Materials Science, 1-2-1 Sengen, Tsukuba, Ibaraki 305-0047, Japan
}%
 \affiliation{Department of Engineering Science, University of Electro-Communications, Chofu, Tokyo 182-8585, Japan}


\date{\today}

\begin{abstract}
This study investigates the quantum correction effect on electrical conductivity using a two-dimensional Wolff Hamiltonian, which is an effective model of the spin-orbit coupling (SOC) lattice system.
The non-adiabatic transition processes in impurity scattering suppress the weak antilocalization (WAL) effect.
The WAL effect in the SOC lattice system strongly depends on the spin relaxation length when compared with the Hikami-Larkin-Nagaoka (HLN) theory.
The spin relaxation length in Bi thin film is discussed.
\end{abstract}
\pacs{Valid PACS appear here}
\maketitle

\section{\label{sec:level1}Introduction}
Systems with strong spin-orbit coupling (SOC) effects have attracted significant attention in the field of spintronics, owing to their potential to generate a large spin current.
However, the spin of these systems is expected to relax quickly, owing to the SOC effect \cite{Elliott1954,Yafet1963,Emoto2016}.
Therefore, it is essential to clarify the criteria to obtain a long spin relaxation length to realize spintronics.
A potential method to evaluate the spin relaxation length is weak localization (WL) analysis using the quantum correction effect \cite{Bergman1982}.
It is well known that the quantum correction effect is described by the Hikami-Larkin-Nagaoka (HLN) theory \cite{HLN1980},
which has been widely used to evaluate the spin relaxation length in systems with strong SOC \cite{Assaf2013,Deorani2014,Peres2014}.

	Crystal atoms that have a strong SOC can be called ``SOC lattices,'' which are different from the case in which impurities have a strong SOC. The SOC lattice system is described by the Hamiltonian equivalent of the Dirac electron system \cite{Wolff1964,Fuseya2015,Hayasaka2016}.
Thus far, the quantum correction effects have been investigated in many Dirac fermion systems such as graphene and surface states of topological insulators \cite{Suzuura2002, McCann2006, ShanWenYu2012, LuHaiZhou2011}. In these Dirac fermion systems, weak antilocalization (WAL) occurs owing to Berry phase $\pi$ effects.
A remarkable feature of an SOC lattice system is its intraband and interband spin hybridization owing to the SOC effect.
The effect of the band-spin hybridization changes the impurity scattering process compared with the case of free electrons.
Even if the impurity potential is diagonal for band and spin indices in the SOC lattice system, the matrix elements between eigenstates with different energies are non-zero elements.
According to Fermi's golden rule, transitions between states with different energies are forbidden by the energy conservation law.
However, in processes that involve higher-order scattering, such as the localization problem, non-adiabatic transitions with different energies are virtually allowed.
The importance of non-adiabatic transitions has been discussed in the context of the anomalous Hall effect \cite{Sinitsyn2007}. 
When non-adiabatic transition processes are included, the understanding of the Berry phase based on the adiabatic picture does not hold, and thus,
it is not obvious whether the WAL completely disappears or partially  remains. In addition, there is a lack of understanding of the difference between the spin relaxation length evaluated by the HLN theory and that based on the Dirac system.\\
\ \ The quantum correction effect is calculated by solving the Bethe-Salpeter equation, which involves correlations between two particles; thus, it has the square of the degrees of freedom of an individual particle. 
When all transition processes are considered in the SOC lattice system, the Bethe-Salpeter equation becomes a ${\rm 16\times 16}$ matrix. Therefore, 16 Cooperons are naively expected to contribute to the quantum correction effect.
In such multiple-degree-of-freedom systems, understanding the experimental results of the quantum correction effects often requires a highly sophisticated interpretation \cite{Eda2016}. Alternatively, it relies on simplification, such as assuming only a single WAL channel by using the HLN formula \cite{Chen2011,Hirahara2014, Akiyama2016}. When a single WAL channel is used, the spin relaxation length is assumed to be sufficiently short. In this case, only the phase relaxation length can be experimentally obtained. \\
\ \ In this study, we consider non-adiabatic transitions in the quantum correction effect using the two-dimensional Wolff Hamiltonian, which is an effective model of the two-dimensional SOC lattice system, such as the {\it L}-point of Bi and PbTe \cite{Wolff1964,Fuseya2015,Hayasaka2016}.
We show that only intraband triplet and interband singlet Cooperons contribute to the quantum correction effect.
By incorporating virtual non-adiabatic transitions, the Cooperon contribution of the interband singlet that leads to WAL is suppressed.
In the weak magnetic field and WAL regimes, we show that the WAL effect increases with an increasing spin relaxation length, in contrast to the HLN theory.
We also show that the quantum correction effect on electrical conductivity clearly depends on the spin relaxation length when compared with the HLN theory.
We demonstrate the WL analysis in Bi thin films and obtain the spin relaxation and phase relaxation lengths.

\section{Model}
For a two-dimensional SOC lattice system, we consider the following Hamiltonian:
\begin{align}
{\cal H}={\cal H}_{0}+V(\bm{r}),
\end{align}
where ${\cal H}_{0}$ is the Wolff Hamiltonian, and $V(\bm{r})$ is the impurity potential. They are given by
\begin{align}
{\cal H}_{0}=\left[
      \begin{array}{cc}
      \Delta  & i\hbar\gamma\bm{\sigma} \cdot \bm{k}\\
      -i\hbar\gamma\bm{\sigma} \cdot \bm{k} &-\Delta\\
      \end{array}
\right],
\end{align}
\begin{align}
V(\bm{r})=u_0\sum_{i}\delta(\bm{r}-\bm{R}_{i}),
\end{align}
where $\hbar$ is Planck's constant, $\bm{k}=(k_{x}, k_{y})$ is the wavenumber vector, $2\Delta$ is the band gap, $\gamma$ is the band parameter, $\sigma$ is the Pauli matrix, $u_{0}$ is the strength of the impurity potential, and $\bm{R}_{i}$ is the impurity position.
The basis of the Wolff Hamiltonian is $\{\ket{c\uparrow},\ket{c\downarrow},\ket{v\uparrow},\ket{v\downarrow}\}$, 
where $\up,\dn$ are the spin degrees of freedom, and $c,v$ are the conduction and valence band degrees of freedom, respectively.
The energy eigenvalues of the Wolff Hamiltonian are $\pm E_{\bm{k}}=\pm\sqrt{\Delta^2+\gamma^2 k^2}$.
For simplicity, we consider the impurity potential to be diagonal for band and spin indices.
The plane wave solutions for $+E_{\bm{k}}$ are
\begin{align}
\ket{1,\bm{k}}
&=
\frac{N_{\bm{k}}}{2}\left(
\begin{array}{c}
1\\
e^{i\phi_{\bm{k}}}\\
-iY\\
-iYe^{i\phi_{\bm{k}}}
\end{array}
\right),
\end{align}
\begin{align}
\ket{2,\bm{k}}
&=
\frac{N_{\bm{k}}}{2}\left(
\begin{array}{c}
-e^{-i\phi_{\bm{k}}}\\
1\\
-iYe^{-i\phi_{\bm{k}}}\\
iY
\end{array}
\right),
\end{align}
where $Y=\hbar\gamma|\bm{k}|/(E_{\bm{k}}+\Delta)$, $N_{\bm{k}}=\sqrt{(\Delta+E_{\bm{k}})/(2E_{\bm{k}})}$, 
$k_x=|\bm{k}|{\rm cos}\phi_{\bm{k}}$, and $k_y=|\bm{k}| {\rm sin}\phi_{\bm{k}}$. The plane wave solutions for $-E_{\bm{k}}$ are  
\begin{align}
\ket{3,\bm{k}}
&=
\frac{N_{\bm{k}}}{2}\left(
\begin{array}{c}
-iY\\
-iYe^{i\phi_{\bm{k}}}\\
1\\
e^{i\phi_{\bm{k}}}
\end{array}
\right),
\end{align}
\begin{align}
\ket{4,\bm{k}}
&=
\frac{N_{\bm{k}}}{2}\left(
\begin{array}{c}
-iYe^{-i\phi_{\bm{k}}}\\
iY\\
-e^{-i\phi_{\bm{k}}}\\
1
\end{array}
\right).
\end{align}

Considering the lowest order Born scattering by the impurity potential, the relaxation time is defined as
\begin{align}
\frac{1}{\tau}&=-2{\rm Im}\Sigma^{\rm R/A}(E_{F})=-2\sum_{\bm{k}'}\average{VG^{\rm R/A}_{0}(\bm{k}')V}_{\rm imp}\nonumber\\
&=2\pi nu_{0}^2\rho_{0},
\end{align}
where $n$ is the impurity concentration, $\rho_{0}$ is the density of states per spin degree of freedom, and $\average{\cdots}_{\rm imp}$ denotes the configuration average of impurities. 
$G_{\alpha0}^{\rm R/A}(\bm{k})=(E_F-E_{\alpha{\bm{k}}}\pm i\delta)^{-1}$ is the non-perturbed single-particle Green's function, where
$E_{1\bm{k}}=E_{2\bm{k}}=E_{\bm{k}}$ and $E_{3\bm{k}}=E_{4\bm{k}}=-E_{\bm{k}}$.
Hereafter, we consider the positive and negative energy eigenstates. The Fermi energy is now sufficiently higher than $\tau^{-1}$, i.e., $E_{F}\tau\gg1$. Even under this condition, the electrons in the negative energy state contribute quantitatively to the quantum correction effect. The case where non-adiabatic transition processes appear is described in Appendix A.
\section{Quantum correction effects on electric and magnetic conductivity}
The effect of quantum correction on the electrical conductivity $\delta\sigma^{(0)}$ (Fig. \ref{Fig1}(a)) is given by \cite{FukuyamaRev1985,Efros_book1985,PALeeRev1985}:
\begin{figure}[t!]
 \begin{center}
  \includegraphics[scale=0.30, bb=150 50 650 680]{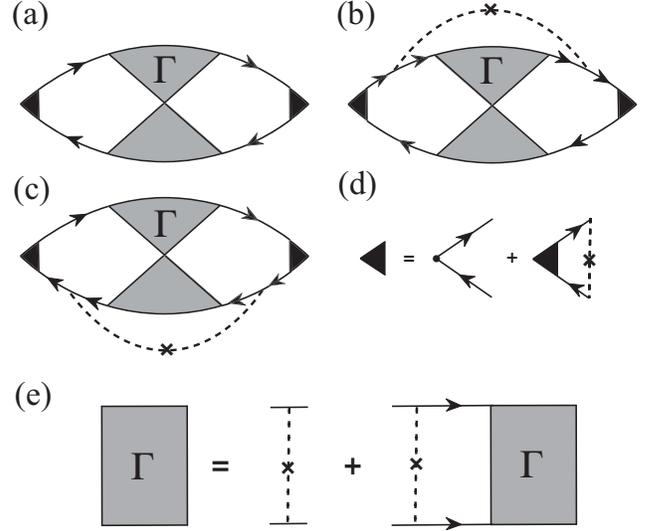}
 \end{center}
 \caption[Quantum correction effect on electrical conductivity.]{Feynman diagrams of the quantum correction effects on electrical conductivity. (a) Quantum corrections of electrical conductivity. (b), (c) Diagrams incorporating the effect of the corrections to the Cooperon. (d), (e) Vertex correction and Bethe-Salpeter equation, respectively.}
  \label{Fig1}
\end{figure}

\begin{align}
\label{kubo}
\delta\sigma^{(0)}(L)&=\frac{e^2\hbar}{2\pi}\sum_{\bm{k}}\tilde{v}^x_{\alpha\omega}(\bm{k})\tilde{v}_{\xi\beta}^{x}(-\bm{k})\nonumber\\
&\times G^{\rm R}_{\alpha}(\bm{k})G^{\rm R}_{\beta}(-\bm{k})G^{\rm A}_{\omega}(\bm{k})G^{\rm A}_{\xi}(-\bm{k})\nonumber\\
&\times \sum_{\bm{q}}\Gamma^{\alpha\beta}_{\xi\omega}(\bm{k},\bm{k},\bm{q}),
\end{align}
where the Einstein summation convention for Greek indices is used.
$G^{\rm R/A}_{\alpha}(\bm{k})=(E_F-E_{\alpha\bm{k}}\pm\rm i\hbar/(2\tau))^{-1}$ denotes the impurity-averaged Green's function.
$\tilde{v}^{x}(\bm{k})$ is the velocity operator with a vertex correction of the form:
\begin{align}
\label{vertex}
&\tilde{v}^{x}_{\beta\alpha}(\bm{k})=v^{x}_{\beta\alpha}+\sum_{\bm{k}'}G^{A}_{\alpha'}(\bm{k}')G^{R}_{\beta'}(\bm{k}')\nonumber\\
&\times \average{\bra{\beta,\bm{k}}V(\bm{r})\ket{\beta',\bm{k}'}
\bra{\alpha', \bm{k}'}V(\bm{r})\ket{\alpha,\bm{k}}}_{\rm imp}\nonumber\\
&\times \tilde{v}^{x}_{\beta'\alpha'}(\bm{k}').
\end{align}
The bare velocity operator $\hat{v^{x}}$ is  given by
\begin{align}
\hat{v^{x}}=\frac{1}{\hbar}\frac{\partial {\cal H}_{0}}{\partial k_{x}}=
\left[
      \begin{array}{cc}
      0  & i\gamma {\sigma_x}\\
      -i\gamma{\sigma_x}  & 0\\
      \end{array}
\right].
\end{align}
Equation (\ref{vertex}) can be solved by assuming a solution of the following form: $\tilde{v}^{x}_{\beta\alpha}=\eta_{v}v^{x}_{\beta\beta}\delta_{\beta\alpha}$,
where $\eta_{v}=2\lambda^2/(\lambda^2+1)$ and $\lambda=E_{\rm F}/\Delta$. 
The quantum correction effect is given by the divergent contribution of $\Gamma^{\alpha\beta}_{\xi\omega}(\bm{k},\bm{k},\bm{q})$ in the particle-particle ladder type scattering process, which causes Cooper instability. 
$\Gamma^{\alpha\beta}_{\xi\omega}(\bm{k},\bm{k},\bm{q})$ is given as a solution to the following Bethe-Salpeter equation \cite{GarateIon2009,GarateIon2012}: 
\begin{align}
\label{eq12}
\Gamma^{mn}_{m'n'}(\bm{q})&=nu_{0}^2\delta_{mn}\delta_{m'n'}\nonumber\\
&+nu_{0}^2\sum_{\bm{k},\ell,\ell'}G^{R}_{lm}(\bm{k})G^{A}_{l'm'}(\bm{q}-\bm{k})\Gamma^{ln}_{l'n'}(\bm{q}).
\end{align}
Here, we change the basis of $\Gamma^{\alpha\beta}_{\xi\omega}$ to $\Gamma^{mn}_{m'n'}(\bm{q})$ using the following relation: 
\begin{align}
\label{ansatz}
\Gamma^{\alpha\beta}_{\xi\omega}(\bm{k}_{1},\bm{k}_2,\bm{q})=\sum_{n,n',m,m'}\braket{\beta,\bm{k}_{2}|n}\braket{\omega,\bm{q}-\bm{k}_{2}|n'}\nonumber\\
\braket{m|\alpha,\bm{k}_{1}}\braket{m'|\xi,\bm{q}-\bm{k}_{1}}\Gamma^{mn}_{m'n'}(\bm{q}).
\end{align}
The basis denoted by $\ket{n}$, $\ket{n'}$, $\ket{m'}$, $\ket{m'}\in \{1,2,3,4\}$ corresponds to $\{\ket{c\up}, \ket{c\dn}, \ket{v\up}, \ket{v\dn}\}$.
$G^{R/A}_{lm}(\bm{k})$ is given by
\begin{align}
\label{eq14}
G^{R/A}_{lm}(\bm{k})=\braket{\alpha,\bm{k}|m}G^{R/A}_{\alpha}(\bm{k})\braket{l|\alpha,\bm{k}}.
\end{align}
Details of the calculations are given in Appendix B.
the components of $\Gamma$ have the following form:
 \begin{align}
\label{5510}
\Gamma^{11}_{11}+\Gamma^{22}_{22}&=\bra{c\up}\otimes\bra{c\up}\hat{\Gamma}\ket{c\up}\otimes\ket{c\up}\nonumber\\
&+\bra{c\dn}\otimes\bra{c\dn}\hat{\Gamma}\ket{c\dn}\otimes\ket{c\dn}\nonumber\\
&=\frac{32nu_{0}^2\lambda^2\pi}{(\lambda^4+\lambda^2+2)v_{F}^2\tau^2q^2+4(\lambda^2-1)},
\end{align}
\begin{align}
\label{5511}
\Gamma^{33}_{33}+\Gamma^{44}_{44}&=\bra{v\up}\otimes\bra{v\up}\hat{\Gamma}\ket{v\up}\otimes\ket{v\up}\nonumber\\
&+\bra{v\dn}\otimes\bra{v\dn}\hat{\Gamma}\ket{v\dn}\otimes\ket{v\dn}\nonumber\\
&=\frac{32nu_{0}^2\lambda^2\pi}{(\lambda^4+\lambda^2+2)v_{F}^2\tau^2q^2+4(\lambda^2-1)},
\end{align}
\begin{align}
\label{5512}
&\Gamma^{11}_{44}-\Gamma^{14}_{41}-\Gamma^{41}_{14}+\Gamma^{44}_{11}\nonumber\\
&=(\bra{c\up}\otimes\bra{v\dn}-\bra{v\dn}\otimes\bra{c\up})\hat{\Gamma}\nonumber\\
&\times(\ket{c\up}\otimes\ket{v\dn}-\ket{v\dn}\otimes\ket{c\up})\nonumber\\
&=\frac{8\pi nu_{0}^2\lambda^2}{(2\lambda^2-1)q^2\tau^2v_{F}^2+2},
\end{align}
\begin{align}
\label{5513}
&\Gamma^{22}_{33}-\Gamma^{23}_{32}-\Gamma^{32}_{23}+\Gamma^{33}_{22}\nonumber\\
&=(\bra{c\dn}\otimes\bra{v\up}-\bra{v\up}\otimes\bra{c\dn})\hat{\Gamma}\nonumber\\
&\times(\ket{c\dn}\otimes\ket{v\up}-\ket{v\up}\otimes\ket{c\dn})\nonumber\\
&=\frac{8\pi nu_{0}^2\lambda^2}{(2\lambda^2-1)q^2\tau^2v_{F}^2+2},
\end{align}
where $v_{F}$ is the Fermi velocity. 
Here, we do not explicitly describe all the components because the other components of $\Gamma$ are zero, or they vanish in the summation in Eq. (\ref{kubo}).
Equations (\ref{5510}) and (\ref{5511}) are intraband triplets, and equations (\ref{5512}) and (\ref{5513}) are interband singlets.
\begin{table*}
 \caption{Components of $\Gamma$ with Cooper instability.}
\label{table:Gamma34}
 \centering
  \begin{ruledtabular}
    \begin{tabular}{clll}
    & Triplet & Singlet& \\
   \hline
   
   SOC lattice &$\ket{c\uparrow}\otimes\ket{c\uparrow}$, $\ket{c\downarrow}\otimes\ket{c\downarrow}, $& $\ket{c\uparrow}\otimes\ket{v\downarrow}-\ket{v\downarrow}\otimes\ket{c\uparrow}$, & \\
   (with non-adiabatic transitions)   &$\ket{v\uparrow}\otimes\ket{v\uparrow}$, $\ket{v\downarrow}\otimes\ket{v\downarrow}$ & $\ket{c\downarrow}\otimes\ket{v\uparrow}-\ket{v\uparrow}\otimes\ket{c\downarrow}$ & \\
   \\
    SOC lattice \cite{Hayasaka2020}  &$\ket{c\uparrow}\otimes\ket{c\uparrow}$, $\ket{c\downarrow}\otimes\ket{c\downarrow}$& $\ket{c\uparrow}\otimes\ket{v\downarrow}-\ket{v\downarrow}\otimes\ket{c\uparrow}$,\\
    (without non-adiabatic transitions)& & $\ket{c\downarrow}\otimes\ket{v\uparrow}-\ket{v\uparrow}\otimes\ket{c\downarrow}$ & \\
     \\   
     HLN \cite{HLN1980} & $\ket{c\uparrow}\otimes\ket{c\uparrow}$, $\ket{c\downarrow}\otimes\ket{c\downarrow}$,  & $\ket{c\uparrow}\otimes\ket{c\downarrow}-\ket{c\downarrow}\otimes\ket{c\uparrow}$ & \\
      ($k_{z}\neq 0$)  &$\ket{c\uparrow}\otimes\ket{c\downarrow}+\ket{c\downarrow}\otimes\ket{c\uparrow}$ &  & \\
   
  \end{tabular}
\end{ruledtabular}
\end{table*}
We obtain the quantum corrections to the electrical conductivity as follows:
\begin{align}
&\delta\sigma^{(0)}(L)=-\frac{e^2}{2\pi^2 \hbar}\sum_{i={cc,vv,s}}\alpha_i {\rm log}\frac{\ell_0^{-2}+\ell_{i}^{-2}}{L^{-2}+\ell_i^{-2}},
\end{align}
where $\alpha_{cc}=(\lambda+1)^2/(\lambda^4+\lambda^2+2)$, $\alpha_{vv}=(\lambda-1)^2/(\lambda^4+\lambda^2+2)$, $\alpha_{s}=-(\lambda^2-1)/[2(2\lambda^2-1)]$, 
$\ell_{cc}^{-2}=\ell_{vv}^{-2}=2(\lambda^2-1)/(\lambda^4+\lambda^2+2)\ell_{0}^{-2}$, $\ell_{s}^{-2}=2/(2\lambda^2-1)\ell_{0}^{-2}$, $\ell_{0}=\sqrt{D_{0}\tau}$, and $D_0=v_{F}^2\tau/2$.
$cc$, $vv$, and $s$ represent conduction-intraband triplets, valence-intraband triplets, and an interband singlet, respectively.
Only intraband triplets and the interband singlet, which is qualitatively equivalent to previous results that do not consider non-adiabatic transitions \cite{Hayasaka2020}, remain.
However, valence-intraband triplets ($\ket{v\up}\otimes\ket{v\up}$ and $\ket{v\dn}\otimes\ket{v\dn}$) also make a divergent contribution.
Table \ref{table:Gamma34} summarizes the components of $\Gamma$ with Cooper instability.
Notably, the intraband triplet ($\ket{c\up}\otimes\ket{c\dn}+\ket{c\dn}\otimes\ket{c\up}$) and intraband singlet ($\ket{c\up}\otimes\ket{c\dn}-\ket{c\dn}\otimes\ket{c\up}$) do not appear, even when transitions between all energy eigenstates are considered.
The absence of the intraband triplet and intraband singlet may be a property that is specific to the exact two-dimensional system with $k_{z}=0$.
Even in the HLN theory with $k_{z}=0$, the intraband singlet and intraband triplet cancel each other out and do not contribute to the quantum correction effect of the electrical conductivity. In the case of the SOC lattice system with $k_{z}=0$, the same cancelation is expected to occur.
It should be noted that in Ref. \cite{GarateIon2012}, the negative energy states are neglected.
If we ignore the negative energy state when solving Eq. (\ref{eq12}) as in Ref. \cite{GarateIon2012}, the result of Ref. \cite{Hayasaka2020} is reproduced.

	Using the lowest-order Cooperon correction, the quantum correction effect on the electrical conductivity (Fig. \ref{Fig1}(b, c)) can be expressed as follows:\begin{align}
\label{556}
&\delta\sigma^{(1)}_{a}=\frac{e^2\hbar}{2\pi}\sum_{\bm{k},\bm{k}_1,\bm{q}}\tilde{v}_{\alpha'\omega}^x(\bm{k})\tilde{v}_{\xi\beta'}^x(-\bm{k}_1)\nonumber\\
&\times G_{\alpha'}^{R}(\bm{k})G_{\alpha}^{R}(\bm{k}_1)G_{\beta}^{R}(-\bm{k})G_{\beta'}^{R}(-\bm{k}_1)\nonumber\\
&\times G_{\xi}^{A}(-\bm{k}_1)G_{\omega}^{A}(\bm{k})
\Gamma^{\alpha\beta}_{\xi\omega}(\bm{q})\nonumber\\
&\average{\braket{\alpha, \bm{k}_1|V(\bm{r})|\alpha',\bm{k}}\braket{\beta',-\bm{k}_1|V(\bm{r})|\beta,-\bm{k}}}_{\rm imp}.
\end{align}
\begin{align}
\label{557}
&\delta\sigma^{(1)}_{b}=\frac{e^2\hbar}{2\pi}\sum_{\bm{k},\bm{k}_1,\bm{q}}\tilde{v}^x_{\alpha\omega'}(\bm{k})\tilde{v}^x_{\xi'\beta}(-\bm{k}_{1})\nonumber\\
&\times G_{\alpha}^{R}(\bm{k})G_{\beta}^{R}(-\bm{k}_1)
G_{\xi'}^{A}(-\bm{k}_1)G_{\xi}^{A}(-\bm{k})\nonumber\\
&\times G_{\omega}^{A}(\bm{k}_1)G_{\omega'}^{A}(\bm{k})
\Gamma^{\alpha\beta}_{\xi\omega}(\bm{q})\nonumber\\
&\times \average{\braket{\xi, -\bm{k}, |V(\bm{r})|\xi', -\bm{k}_1}\braket{\omega', \bm{k}|V(\bm{r})|\omega, \bm{k}_1}}_{\rm imp}.
\end{align}
The quantum correction $\delta\sigma_{\rm W}$ to the electrical conductivity, considering all contributions from Fig. 1(a), (b), and (c), becomes
\begin{align}
\delta\sigma_{\rm W}&=\delta\sigma^{(0)}+\delta\sigma^{(1)}_a+\delta\sigma^{(1)}_b\nonumber\\
&=-\frac{e^2}{2\pi^2\hbar}\eta_{v}^2\sum_{i=cc,vv,s}(1+2\eta_{H,i})\alpha_i {\rm log}\frac{\ell_0^{-2}+\ell_{i}^{-2}}{L^{-2}+\ell_i^{-2}}.
\end{align}
where $\eta_{H,cc}=-(\lambda-1)/(4\lambda)$, $\eta_{H,vv}=-(\lambda+1)/(4\lambda^2)$ and $\eta_{H,s}=-1/4$.

\begin{figure}[t!]
 \begin{center}
  \includegraphics[scale=0.33, bb=150 50 650 550]{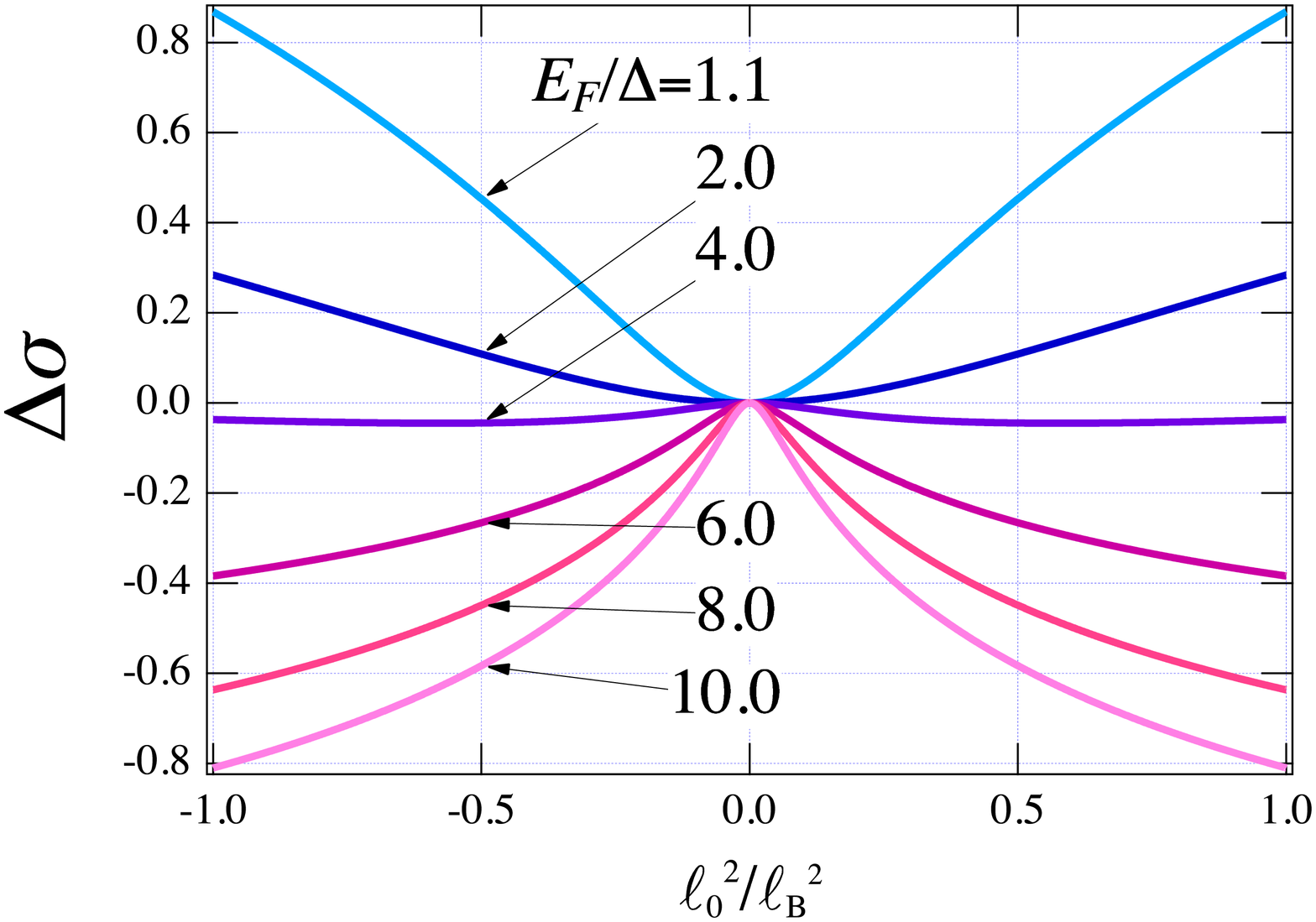}
 \end{center}
 \caption[Magnetic field dependence of $\delta \sigma_{\rm W}$]{Magnetic field dependence of $\Delta\sigma(B)=[\delta\sigma_{\rm W}(B)-\delta\sigma_{\rm W}(0)]/(e^2/2\pi^2\hbar)$. }
 \label{fig:MC}
\end{figure}

The electrical conductivity in a magnetic field $\delta\sigma_{\rm W}(B)=\delta\sigma^{(0)}(B)+\delta\sigma^{(1)}_a(B)+\delta\sigma^{(1)}_b(B)$ is obtained as follows:
\begin{align}
\label{eqMC}
\delta&\sigma_{\rm W}(B)=-\frac{e^2}{2\pi^2\hbar}\eta_{v}^2\sum_{i=cc,vv,s}\alpha_{i}(1+2\eta_{H,i})\nonumber\\
&\times\Biggl[\psi\left(\frac{1}{2}+\frac{\ell_{B}^2}{\ell_{0}^2}+\frac{\ell_{B}^2}{\ell_i^2}+\frac{\ell_{B}^2}{\ell_{\phi}^2}\right)-\psi\left(\frac{1}{2}+\frac{\ell_{B}^2}{\ell_i^2}+\frac{\ell_{B}^2}{\ell_{\phi}^2}\right)\Biggr],
\end{align}
where $\psi$ is the digamma function and $\ell_{B}=\sqrt{\hbar/4eB}$ is the magnetic length of the electron pair.
The magnetic field dependence of the quantum correction effect on the electrical conductivity is plotted in Fig. \ref{fig:MC}. By changing $E_{\rm F}/\Delta$, a WL-WAL crossover occurs. 


\begin{figure}[t!]
\begin{center}
  \includegraphics[scale=0.33, bb=150 50 650 550]{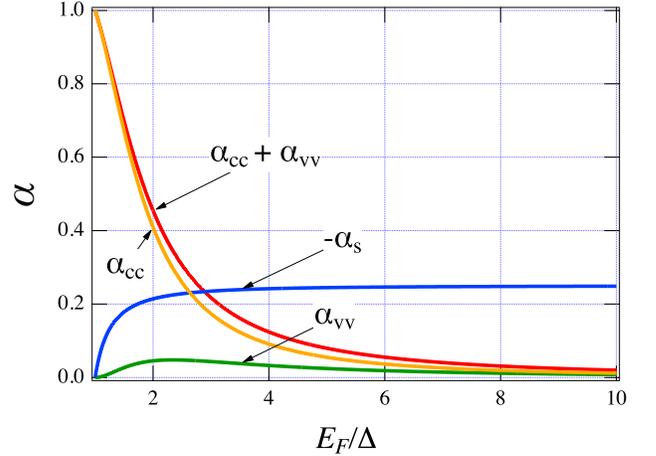}
 \end{center}
 \caption{Dependence of $\alpha_{cc}$, $\alpha_{vv}$, and -$\alpha_{s}$ on $E_{F}/\Delta$. }
\label{alpha5}
\end{figure}
The WL-WAL crossover is characterized by the strength of the intraband triplet $\alpha_{cc}+\alpha_{vv}$ and interband singlet $\alpha_{s}$. Figure 3 shows the dependence of $\alpha$ on $E_{\rm F}/\Delta$.
The conduction-intraband triplet $\alpha_{cc}$ and interband singlet $\alpha_{s}$ dominate for $E_{\rm F}/\Delta\sim 1$ and $E_{\rm F}/\Delta\rightarrow\infty$, respectively.
The valence-intraband triplet $\alpha_{vv}$ exhibits a gentle peak in the intermediate $E_{\rm F}/\Delta$ region. 
Consequently, the WL-WAL crossover occurs at $E_{F}/\Delta \sim 3$. This value agrees with previous results \cite{Hayasaka2020}.
The interband singlet $\alpha_{s}$ becomes $1/4$ as $ E_{\rm F}/\Delta \rightarrow \infty$. This value is smaller than $\alpha_{s}=1/2(E_{\rm F}/\Delta \rightarrow \infty)$ when only adiabatic processes are considered \cite{Hayasaka2020}; thus, the WAL effect is suppressed.
This can be intuitively understood from the matrix elements of impurity scattering.
During the transition of an electron from the $\ket{1,\bm{k}}$ state to the $\ket{1,\bm{k}'}$ state, owing to impurities, 
the matrix elements of the impurity scattering become
\begin{align}
\label{20}
\braket{1,\bm{k}'|V(\bm{r})|1,\bm{k}}\propto(1+Y^2)(1+e^{i(\phi-\phi')}).
\end{align}
The backscattering process can be obtained by making the following substitution: $\bm{k}'\rightarrow-\bm{k}$, that is, $\phi'\rightarrow\phi+\pi$; thus, equation (\ref{20}) becomes zero. This is similar to what is well known for graphene and the surface states of topological insulators, which indicates that the backscattering process is suppressed \cite{Ando1998, Suzuura2002, LuHaiZhou2011, ShanWenYu2012}. Thus, this process contributes to the WAL.
For the transition from $\ket{1,\bm{k}}$ to $\ket{2,\bm{k}'}$, the following relationship holds:
\begin{align}
\label{21}
\braket{2,\bm{k}'|V(\bm{r})|1,\bm{k}}\propto(1-Y^2)(e^{i\phi'}-e^{i\phi}).
\end{align}
For  $E_{\rm F}/\Delta\rightarrow\infty$, $Y\rightarrow 1$; thus, this process also contributes to the WAL in the large $E_{\rm F}/\Delta$ region.
In contrast, the non-adiabatic transition process from $\ket{1,\bm{k}}$ to $\ket{4,\bm{k}'}$ becomes
\begin{align}
\label{23}
\braket{4,\bm{k}'|V(\bm{r})|1,\bm{k}}\propto 2iYe^{i\phi'}-2iYe^{i\phi}.
\end{align}
Equation (\ref{23}) does not contribute to the WAL as an adiabatic process (Equations (\ref{20}) and (\ref{21})).
Therefore, when this process is considered, the sum of all transition probabilities is conserved, which weakens the effect of WAL compared with the case for which this process is not considered. 

\section{Spin relaxation length}
\begin{figure}[t!]
 \begin{center}
  \includegraphics[scale=0.33, bb=150 50 650 550]{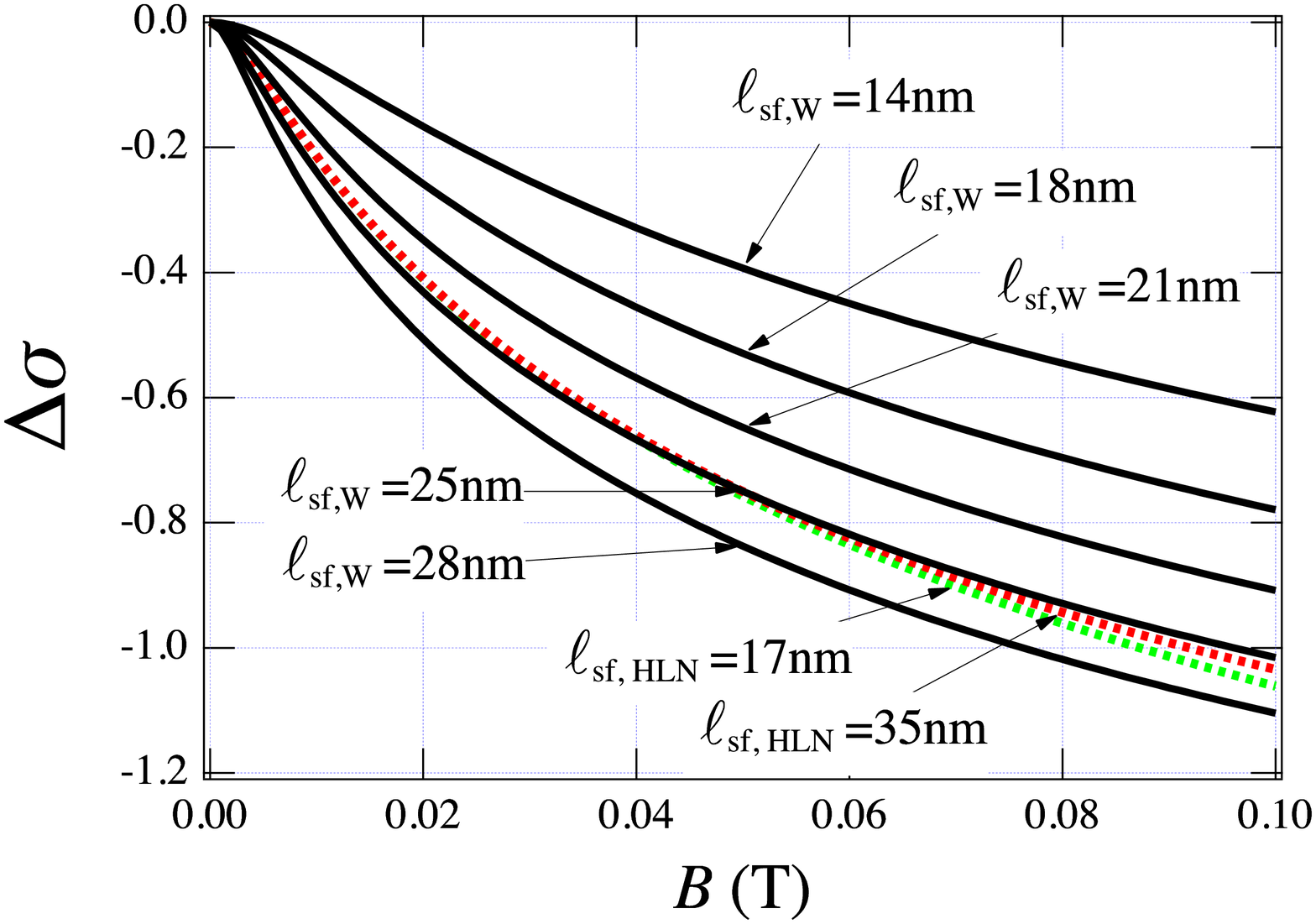}
 \end{center}
\caption[Comparison of $\delta\sigma_{\rm W}$ and $\delta\sigma_{\rm HLN}$]{Dependence of $\Delta\sigma(B)=[\delta\sigma(B)-\delta\sigma(0)]/(e^2/2\pi^2\hbar)$ on the magnetic field. The solid lines correspond to $\delta\sigma_{\rm W}$, and the dotted line corresponds to $\delta\sigma_{\rm HLN}$. In $\delta\sigma_{\rm W}$, $E_{F}/\Delta=10$, $\ell_{\phi}=1000{\rm nm}$, and ${\ell_{0}=20,25,30,35,40{\rm nm}}$. We used $\ell_{{\rm sf},{\rm W}}\sim\ell_{0}/\sqrt{2}$. For $\delta\sigma_{\rm HLN}$, we set $\ell_{\phi}=300{\rm nm}$, $\ell_{0}=20{\rm nm}$,  $\ell_{\rm so}=20,40{\rm nm}$, and we used $\ell_{{\rm sf},{\rm HLN}}=\sqrt{3}/2\ell_{\rm so}$.}
 \label{WvsH}
\end{figure}
In this section, we compare the evaluation of the spin relaxation length with that of the HLN theory.
The spin relaxation length can be estimated by fitting the formula for the quantum correction effect on the electrical conductivity in the weak-field region ($\ell_{B}\gg\ell_{0}$).
In the SOC lattice system, the coupling of spin and momentum results in a simultaneous relaxation of the spin with the relaxation of the momentum.
Therefore, the spin relaxation length is related to $\ell_{0}$, and it can be evaluated as $\ell_{{\rm sf},{\rm W}}\sim\ell_{0}/\sqrt{2}$.
$\sqrt{2}$ arises from doubling of the spin. The HLN formula is given by \cite{HLN1980}:

\begin{align}
\label{NormalHLN}
&\delta\sigma_{\rm HLN}(B)=-\frac{e^2}{2\pi^2\hbar}\Biggl[\frac{3}{2}\Biggl\{\Psi\Bigg(\frac{1}{2}+\frac{\ell_B^2}{\ell_0^2}+\frac{\ell_B^2}{\ell_{\rm so}^2}+\frac{\ell_B^2}{\ell_{\phi}^2}\Biggr)\nonumber\\
&-\Psi\Bigg(\frac{1}{2}+\frac{4}{3}\frac{\ell_B^2}{\ell_{\rm so}^2}+\frac{\ell_B^2}{\ell_{\phi}^2}\Biggr)\Biggr\}
-\frac{1}{2}\Biggl\{\Psi\Biggl(\frac{1}{2}+\frac{\ell_B^2}{\ell_{0}^2}+\frac{\ell_B^2}{\ell_{\rm so}^2}+\frac{\ell_B^2}{\ell_{\phi}^2}\Biggr)\nonumber\\
&-\Psi\Biggl(\frac{1}{2}+\frac{\ell_B^2}{\ell_{\phi}^2}\Biggr)\Biggr\}\Biggr].
\end{align}
The relation between the spin relaxation time and the spin-orbit relaxation time is given by $1/\tau_{s}=4/(3\tau_{\rm so})$ \cite{DasSarma2004}. 
Therefore, the spin relaxation length in the HLN theory is given by $\ell_{{\rm sf},{\rm HLN}}=\sqrt{3}/2\ell_{\rm so}$.
Figure 4 presents a plot of $\delta\sigma_{\rm W}$ and $\delta\sigma_{\rm HLN}$.
Compared with $\delta\sigma_{\rm HLN}$, $\delta\sigma_{\rm W}$ shows a sharp change in the WAL effect with respect to the change in the spin relaxation length.
This is because the instability of the interband singlet, which leads to the WAL effect, is essentially inseparable from the SOC effect.
In fact, the interband singlet contains information about the spin relaxation through $\ell_{0}$.
In the case of the HLN theory, the intraband singlet is not affected by the SOC effect. As a result, the intraband singlet does not contain the spin relaxation length, but only the phase relaxation length $\ell_\phi$.
Therefore, if $\ell_\phi$ is constant, $\delta\sigma_{\rm HLN}$ slightly changes in the weak-field region with respect to the change in the spin relaxation length.

A conventional WL analysis using the HLN theory suggests that the WAL effect can be observed only when the spin relaxation length is sufficiently short.
Furthermore, as previously mentioned, the conductivity of the HLN theory is almost independent of the spin relaxation length in the weak magnetic field and the WAL regime; hence, the quantum correction effect was analyzed using only the intraband singlet \cite{Chen2011,Hirahara2014, Akiyama2016}. In this analysis, only the phase relaxation length can be obtained. However, the WAL effect in the SOC lattice system is enhanced by an increasing spin relaxation length, and it is more sensitive to changes in the spin relaxation length than $\delta\sigma_{\rm HLN}$.
Therefore, $\delta\sigma_{\rm W}$ can extract more rich information than $\delta\sigma_{\rm HLN}$.

\section{WL analysis}
\begin{figure}[t!]
 \begin{center}
  \includegraphics[scale=0.32, bb=150 50 720 600]{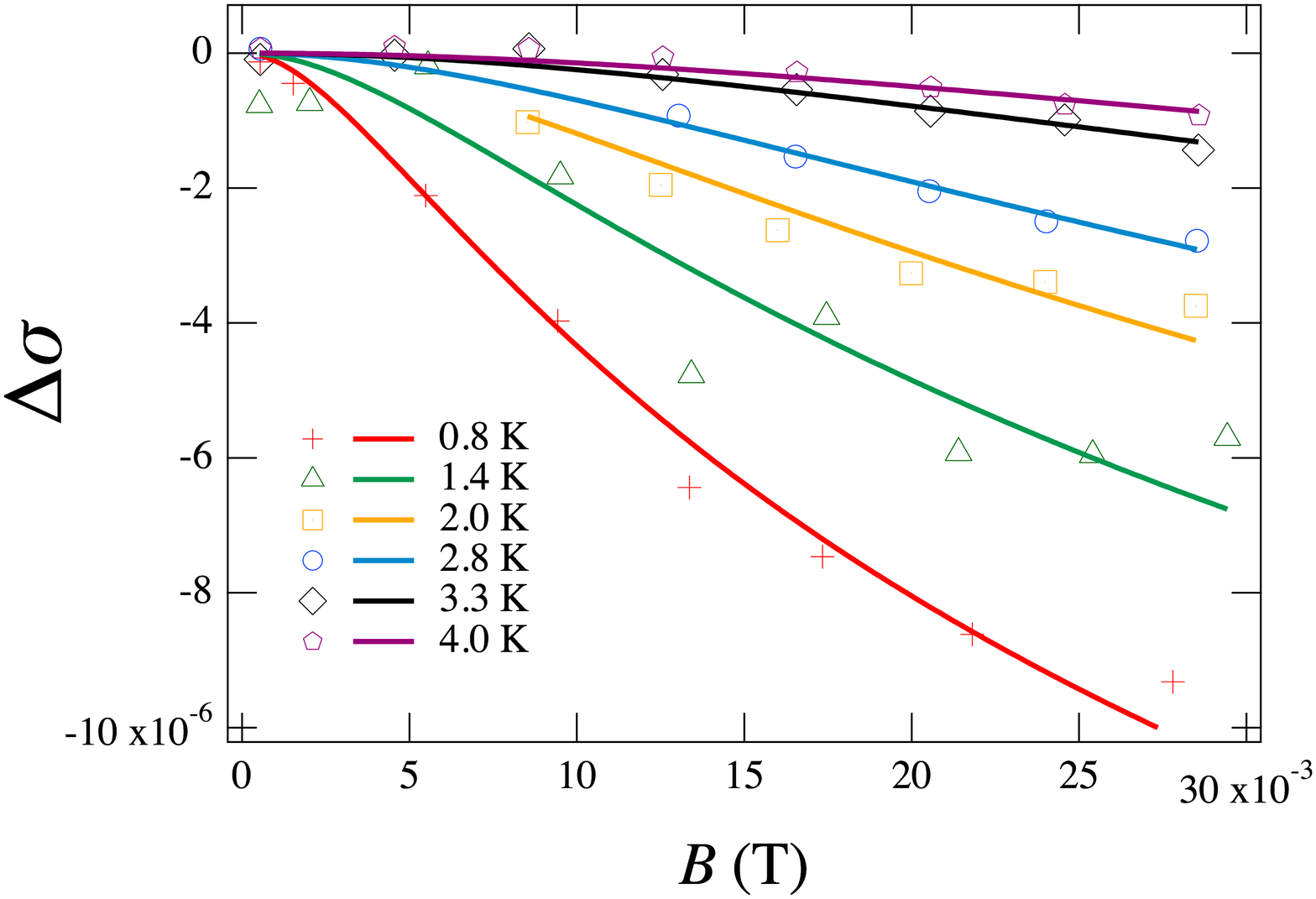}
 \end{center}
\caption[Fitting of experimental data]{The temperature dependence of conductivity in the WAL regime for a 20 bilayer (BL) Bi film (These values are extracted from Ref. \cite{Hirahara2014}). Solid lines represent $3\times( \delta\sigma_{\rm W}(B)-\delta\sigma_{\rm W}(0))$. }
 \label{WLanalysis}
\end{figure}
Finally, we perform the WL analysis for Bi thin films . According to the recent experiments of Aitani et al., the transport properties of 20 bilayer (BL) Bi thin films are dominated by bulk properties rather than surface states \cite{Hirahara2014}. We focus on the 20 BLs to ignore the effect of the surface states.
The Fermi energy and band gap are given by $E_{F}=35.3\ {\rm meV}$ and $\Delta=7.7\ {\rm meV}$, respectively \cite{Fuseya2012}. Therefore, $\lambda=4.6$.
As Bi has three equivalent electron surfaces \cite{Kamran2011}, we can assume that these contributions are additive and multiply the coefficient of $\delta \sigma_{\rm W}$ by three.
In the fitting procedure, a somewhat technical method is used owing to the complex parameter space resulting from the nonlinear functions \cite{Peres2014}.
We use the following procedure  to reduce this difficulty: We assume that the phase relaxation length is infinite and determine the spin relaxation length at 0.8 K. As impurity scattering rather than electron-lattice scattering is dominant at low temperatures, the phase relaxation length is used as a fitting parameter for 1.4 K--4.0 K and used as the fixed value of the spin relaxation length obtained at 0.8 K.
Note that in Ref. \cite{Hirahara2014}, although the magnet conductivity is studied up to 0.2 T, the classical contribution $\propto B^2$ overlaps with the diffusive contribution at a higher magnetic field \cite{Assaf2013}.
To neglect this contribution, we limited the range up to 0.03 T in the fitting procedure.
The results of the fitting and the fitting parameters are shown in Fig. 5 and Table \ref{table2}, respectively. 
In this manner, the spin relaxation length can be determined as $\ell_{\rm sf,W}=44.9\ {\rm nm}$ in the 20 BL Bi film.
\begin{table}
 \caption{spin relaxation and phase relaxation lengths.}
\label{table2}
 \centering
  \begin{ruledtabular}
    \begin{tabular}{clll}
    & $\ell_{\rm sf,W}$ & $\ell_{\phi}$ & \\
   \hline
     0.8 K  & 44.9 nm & $ \infty $ & \\
     1.4 K & & 336 nm   & \\
     2.0 K & & 300 nm & \\
     2.8 K & & 183 nm & \\
     3.3 K & & 129 nm & \\
     4.0 K & & 112 nm & \\
  \end{tabular}
\end{ruledtabular}
\end{table}
\section{Conclusion}

We investigated the quantum correction effect based on an effective model of the SOC lattice system.
We showed that the WAL effect is suppressed when non-adiabatic transitions are considered, compared with the case in which only adiabatic transitions are considered.
We found that only intraband triplets and interband spin singlets contribute to Cooper instability, even if non-adiabatic transitions are included.
This significantly simplifies the interpretation of the experimental results of the quantum correction effects in SOC lattice systems.
The WAL effect in the SOC lattice system is sensitive to changes in the spin relaxation length, and it increases for longer spin relaxation lengths in contrast to the HLN theory. We expect that our results on the quantum correction effect will be useful for the quantitative evaluation of the spin relaxation length in SOC lattice systems.

\begin{acknowledgments}
The author would like to thank Y. Fuseya for commenting on the manuscript. 
\end{acknowledgments}
\appendix
\section{Non-adiabatic transition}
When $E_{F}\tau \gg 1$ is satisfied, the integral involving the product of Green's functions with different energies is a negligibly small quantity.
However, when virtual non-adiabatic transitions occur owing to impurity scattering, certain diagrams give non-negligible contributions.
It is possible to confirm this by considering the lowest-order crossed diagrams $\Gamma_{\rm a}^{(2)}$ and $\Gamma_{\rm b}^{(2)}$ (see Fig. 6). 
$\Gamma_{\rm a}^{(2)}$ contains the adiabatic transition, and $\Gamma_{\rm b}^{(2)}$ contains the non-adiabatic transition.
In this calculation, we can set $\bm{q}=0$ without loss of generality.
$\Gamma_{\rm a}^{(2)}$ is given by
\begin{align}
\Gamma_{\rm a}^{(2)}&=\sum_{\bm{k}'}\average{\braket{1,-\bm{k}'|V(\bm{r})|1,-\bm{k}}\braket{1,\bm{k}'|V(\bm{r})|1,\bm{k}}}_{\rm imp}\nonumber\\
&\times\average{\braket{1,\bm{k}|V(\bm{r})|1,-\bm{k}'}\braket{1,-\bm{k}|V(\bm{r})|1,\bm{k}'}}_{\rm imp}\nonumber\\
&\times G^{R}_{1}(\bm{k}')G^{A}_{1}(-\bm{k}').
\label{eq1}
\end{align}
$\Gamma_{\rm b}^{(2)}$ is given by
\begin{align}
\Gamma_{\rm b}^{(2)}&=\sum_{\bm{k}'}\average{\braket{4,-\bm{k}'|V(\bm{r})|1,-\bm{k}}\braket{4,\bm{k}'|V(\bm{r})|1,\bm{k}}}_{\rm imp}\nonumber\\
&\times\average{\braket{1,\bm{k}|V(\bm{r})|4,-\bm{k}'}\braket{1,-\bm{k}|V(\bm{r})|4,\bm{k}'}}_{\rm imp}\nonumber\\
&\times G^{R}_{4}(\bm{k}')G^{A}_{4}(-\bm{k}')
\label{eq2}
\end{align}
These can be calculated as follows:
\begin{align}
\Gamma_{\rm a}^{(2)}=\frac{n^2u_{0}^4}{4}\pi \rho_{0}\tau N^8 (1+Y^2)^{4},
\label{eq3}
\end{align}
\begin{align}
\Gamma_{\rm b}^{(2)}=-4n^2u_{0}^4\pi \rho_{0}\tau N^8Y^4.
\label{eq4}
\end{align}
Therefore, Eq. (\ref{eq4}) cannot be neglected even under the condition $E_F \tau\gg1$ and gives the same order as Eq. (\ref{eq3}).
In fact, many studies \cite{Suzuura2002, ShanWenYu2012, LuHaiZhou2011, GarateIon2012} neglect the process described in Eq. (\ref{eq2}) and our previous calculation \cite{Hayasaka2020} also follows this approach.
Even with such an approximation, the qualitative features of the quantum correction effect can be captured effectively. 
However, to quantitatively evaluate the spin relaxation length, all the transition processes should be included.
\begin{figure}[t!]
 \begin{center}
  \includegraphics[scale=0.50, bb=150 50 350 250]{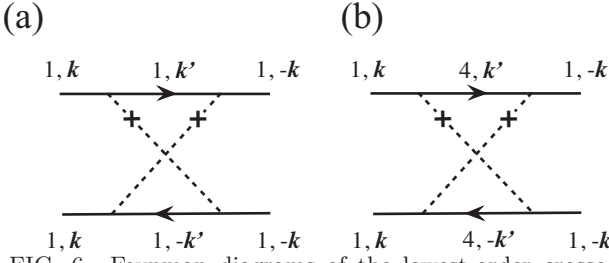}
 \end{center}
 \caption[Quantum correction effect on electrical conductivity.]{Feynman diagrams of the lowest-order crossed diagram: (a) adiabatic and (b) non-adiabatic transition processes. }
  \label{fig:sigma55}
\end{figure}

\section{Bethe-Salpeter equation}
The main task in the calculation of quantum correction effects is to solve the Bethe-Salpeter equation.
This is an elementary but tedious calculation.
Equation (\ref{eq12}) can be expressed in the matrix product form as follows:
\begin{align}
\label{eq28}
(\hat{\rm I}-nu_{0}^2\hat{\Pi})\hat{\Gamma}=nu_{0}^2\hat{\rm I}, 
\end{align}
where ${\rm \hat{I}}$ is the $16\times 16$ unit matrix.
$\Pi^{lm}_{l'm'}$ is given by
\begin{align}
\Pi^{lm}_{l'm'}=\Big\langle \rho_{0}\int dE_{\bm{k}}G^{R}_{lm}(\bm{k})G^{A}_{l'm'}(\bm{q}-\bm{k})\Big\rangle_{\rm F}.
\label{eqB2}
\end{align}
$\average{\cdots}_{F}$ denotes the angle average on the Fermi surface. Here, as the momentum on the Fermi surface is relevant, the summation of the wavenumbers is replaced by the following:
\begin{align}
\label{537}
\sum_{\bm{k}}\rightarrow \Big\langle\rho_{0}\int dE_{\bm{k}}\Big\rangle_{\rm F}.
\end{align}
From equation (\ref{eq14}), $G^{R}_{lm}$ and $G^{A}_{l'm'}$ can be expressed as follows:
\begin{align}
&G^{R}_{11}(\bm{k})=G^{R}_{22}(\bm{k})=G^{R}_{1}(\bm{k})N^2+G^{R}_{4}(\bm{k})N^2Y^2,\\
&G^{R}_{14}(\bm{k})=G^{R\ast}_{41}(\bm{k})\nonumber\\
&=iG^{R}_{1}(\bm{k})N^2Y e^{-i\phi_{\bm{k}}}-iG^{R}_{4}(\bm{k})N^2Ye^{-i\phi_{\bm{k}}},\\
&G^{R}_{23}(\bm{k})=G^{R\ast}_{32}(\bm{k})\nonumber\\
&= iG^{R}_{1}(\bm{k})N^2Ye^{i\phi_{\bm{k}}}-iG^{R}_{4}(\bm{k})N^2Ye^{i\phi_{\bm{k}}},\\
&G^{R}_{33}(\bm{k})=G^{R}_{44}(\bm{k})=G^{R}_{1}(\bm{k})N^2Y^2+G^{R}_{4}(\bm{k})N^2,
\end{align}
\begin{align}
&G^{A}_{11}(\bm{q}-\bm{k})=G^{A}_{22}(\bm{q}-\bm{k})\nonumber\\
&=G^{A}_{1}(\bm{q}-\bm{k})N^2+G^{A}_{4}(\bm{q}-\bm{k})N^2Y^2,\\
&G^{A}_{14}(\bm{q}-\bm{k})=G^{A\ast}_{41}(\bm{q}-\bm{k})\nonumber\\
&=-iG^{A}_{1}(\bm{q}-\bm{k})N^2Y e^{-i\phi_{\bm{k}}}+iG^{A}_{4}(\bm{q}-\bm{k})N^2Ye^{-i\phi_{\bm{k}}},\\
&G^{A}_{23}(\bm{q}-\bm{k})=G^{A\ast}_{32}(\bm{q}-\bm{k})\nonumber\\
&= -iG^{A}_{1}(\bm{q}-\bm{k})N^2Ye^{i\phi_{\bm{k}}}+iG^{A}_{4}(\bm{q}-\bm{k})N^2Ye^{i\phi_{\bm{k}}},\\
&G^{A}_{33}(\bm{q}-\bm{k})=G^{A}_{44}(\bm{q}-\bm{k})\nonumber\\
&=G^{A}_{1}(\bm{q}-\bm{k})N^2Y^2+G^{A}_{4}(\bm{q}-\bm{k})N^2,
\end{align}
and the other components of $G^{R}_{lm}$ and $G^{A}_{l'm'}$ are zero.
The components of $\Pi^{lm}_{l'm'}$ are as follows:
For $(l,m)=(1,1)$ and $(l,m)=(2,2)$,
\begin{align}
\Pi^{lm}_{11}&=\Pi^{lm}_{22}=N^4\pi\rho_{0}\tau(1+Y^4)(2-Q^2),\\
\Pi^{lm}_{14}&=N^4\pi\rho_{0}\tau Y(1+Y^2)(Q_{x}+iQ_{y}),\\
\Pi^{lm}_{23}&=N^4\pi\rho_{0}\tau Y(1+Y^2)(Q_{x}-iQ_{y}),\\
\Pi^{lm}_{32}&=-N^4\pi\rho_{0}\tau Y(1+Y^2)(Q_{x}+iQ_{y}),\\
\Pi^{lm}_{33}&=\Pi^{lm}_{44}=N^4\pi\rho_{0}\tau Y^2 2(2-Q^2),\\
\Pi^{lm}_{41}&=N^4\pi\rho_{0}\tau Y(1+Y^2)(-Q_{x}+iQ_{y}); 
\end{align}
for $(l,m)=(3,3)$ and $(l,m)=(4,4)$, 
\begin{align}
\Pi^{lm}_{11}&=\Pi^{lm}_{22}=N^4\pi\rho_{0}\tau Y^22(2-Q^2),\\
\Pi^{lm}_{14}&=N^4\pi\rho_{0}\tau Y(1+Y^2)(Q_{x}+iQ_{y}),\\
\Pi^{lm}_{23}&=N^4\pi\rho_{0}\tau Y(1+Y^2)(Q_{x}-iQ_{y}),\\
\Pi^{lm}_{32}&=-N^4\pi\rho_{0}\tau Y(1+Y^2)(Q_{x}+iQ_{y}),\\
\Pi^{lm}_{33}&=\Pi^{lm}_{44}=N^4\pi\rho_{0}\tau (1+Y^4)(2-Q^2),\\
\Pi^{lm}_{41}&=N^4\pi\rho_{0}\tau Y(1+Y^2)(-Q_{x}+iQ_{y}); 
\end{align}
for $(l,m)=(1,4)$, 
\begin{align}
\Pi^{14}_{11}&=\Pi^{14}_{22}=-N^4\pi\rho_{0}\tau Y (1+Y^2)(Q_{x}+iQ_{y}),\\
\Pi^{14}_{14}&=-\Pi^{14}_{32}=-N^4\pi\rho_{0}\tau Y^2 (Q_{x}+iQ_{y})^2,\\
\Pi^{14}_{41}&=-\Pi^{14}_{23}=-N^4\pi\rho_{0}\tau Y^2 2(2-Q^2),\\
\Pi^{14}_{44}&=\Pi^{14}_{33}=-N^4\pi\rho_{0}\tau Y (1+Y^2)(Q_{x}+iQ_{y});
\end{align}
for $(l,m)=(4,1)$,
\begin{align}
\Pi^{41}_{11}&=\Pi^{41}_{22}=N^4\pi\rho_{0}\tau iY (1+Y^2)(Q_{x}-iQ_{y}),\\
\Pi^{41}_{14}&=-\Pi^{41}_{32}=-N^4\pi\rho_{0}\tau Y^2 2(2-Q^2),\\
\Pi^{41}_{41}&=-\Pi^{41}_{23}=-N^4\pi\rho_{0}\tau Y^2 (Q_{x}-iQ_{y})^2,\\
\Pi^{41}_{44}&=\Pi^{41}_{33}=N^4\pi\rho_{0}\tau Y (1+Y^2)(Q_{x}-iQ_{y});
\end{align}
for $(l,m)=(2,3)$,
\begin{align}
\Pi^{23}_{11}&=\Pi^{23}_{22}=-N^4\pi\rho_{0}\tau Y (1+Y^2)(Q_{x}-iQ_{y}),\\
\Pi^{23}_{14}&=-\Pi^{23}_{32}=N^4\pi\rho_{0}\tau Y^2 2(2-Q^2),\\
\Pi^{23}_{41}&=-\Pi^{23}_{23}=N^4\pi\rho_{0}\tau Y^2 \pi(Q_{x}-iQ_{y})^2,\\
\Pi^{23}_{44}&=\Pi^{23}_{33}=N^4\pi\rho_{0}\tau Y (1+Y^2)(-Q_{x}+iQ_{y});
\end{align}
for $(l,m)=(3,2)$,
\begin{align}
\Pi^{32}_{11}&=\Pi^{32}_{22}=N^4\pi\rho_{0}\tau Y (1+Y^2)(Q_{x}+iQ_{y}),\\
\Pi^{32}_{14}&=-\Pi^{32}_{32}=N^4\pi\rho_{0}\tau Y^2 \pi(Q_{x}+iQ_{y})^2,\\
\Pi^{32}_{41}&=-\Pi^{32}_{23}=N^4\pi\rho_{0}\tau Y^2 2(2-Q^2),\\
\Pi^{32}_{44}&=\Pi^{32}_{33}=N^4\pi\rho_{0}\tau Y (1+Y^2)(Q_{x}+iQ_{y});
\end{align}
the other components of $\Pi^{lm}_{l'm'}$ are zero, where we define $Q_{x}=(\gamma^2|\bm{k}_{F}|/E_{F})\tau q_{x}$, $Q_{y}=(\gamma^2|\bm{k}_{F}|/E_{F})\tau q_{y}$.
After solving for $\Gamma$ and leaving it to the order of $q^2$, the diagonal components of $\Gamma$ are as follows:
\begin{align}
\Gamma^{11}_{11}=\frac{16\lambda^2\pi nu_{0}^2}{(\lambda^4+\lambda^2+2)q^2\tau^2v_F^2+4(\lambda^2-1)},
\end{align}
\begin{align}
\Gamma^{11}_{22}&=\frac{(\lambda^2+1)\pi nu_{0}^2}{\lambda^2q^2\tau^2v_F^2},
\end{align}
\begin{align}
\Gamma^{11}_{33}&=\frac{8\lambda^2(3\lambda^2+1)^2\pi nu_{0}^2}{(7\lambda^6-\lambda^4-3\lambda^2-3)q^2\tau^2v_F^2+24\lambda^6+32\lambda^4+8\lambda^2},
\end{align}
\begin{align}
\Gamma^{11}_{44}&=\frac{2(\lambda^2+1)^2\pi nu_{0}^2}{(2\lambda^4-\lambda^2)q^2\tau^2v_F^2+2\lambda^2+2}.
\end{align}
The other diagonal components are given by
$\Gamma^{11}_{11}=\Gamma^{22}_{22}=\Gamma^{33}_{33}=\Gamma^{44}_{44}$, 
$\Gamma^{11}_{22}=\Gamma^{22}_{11}=\Gamma^{33}_{44}=\Gamma^{44}_{33}$, 
$\Gamma^{11}_{33}=\Gamma^{22}_{44}=\Gamma^{33}_{11}=\Gamma^{44}_{22}$ and 
$\Gamma^{11}_{44}=\Gamma^{22}_{33}=\Gamma^{33}_{22}=\Gamma^{44}_{11}$.
In the non-diagonal components, the non-zero terms after the angular integration of $\bm{q}$ are
\begin{align}
\Gamma^{14}_{41}=-\frac{2(\lambda^2-1)^2\pi nu_{0}^2}{(2\lambda^4-\lambda^2)q^2\tau^2v_{F}^2+2(\lambda^2-1)}, 
\end{align}
$\Gamma^{23}_{32}=\Gamma^{32}_{23}=\Gamma^{41}_{14}=\Gamma^{14}_{41}$, 
$\Gamma^{14}_{23}=\Gamma^{23}_{14}=\Gamma^{32}_{41}=\Gamma^{41}_{32}=\Gamma^{11}_{22}$,  
and the other components of $\Gamma^{lm}_{l'm'}$ are zero. It should be noted that $\Gamma^{11}_{22}$ ( $\Gamma^{14}_{23}, \Gamma^{23}_{14}, \Gamma^{32}_{41}$, and $\Gamma^{41}_{32}$) do not have the $q^{0}$ term. 
However, these gapless terms vanish in the summation of Eq. (9).
\bibliography{apssamp2}
\end{document}